\documentclass{llncs}
\usepackage{llncsdoc}
\usepackage{graphicx} % Allows including images
\usepackage{booktabs} % Allows the use of \toprule, \midrule and \bottomrule in tables
\usepackage{amsmath}
\usepackage{amsfonts}
\usepackage{amssymb}
\usepackage{epstopdf}
\usepackage{dblfloatfix}
\usepackage[caption=false,font=footnotesize]{subfig}
\usepackage{color}
\usepackage{enumitem}
\begin{document}

\newcounter{save}\setcounter{save}{\value{section}}
{\def\addtocontents#1#2{}%
\def\addcontentsline#1#2#3{}%
\def\markboth#1#2{}%
\title{Modemless Multiple Access Communications over Powerlines for DC Microgrid Control}

\author{Marko Angjelichinoski, \v Cedomir Stefanovi\' c, Petar Popovski} %Marko, \v Cedomir, Petar}

\institute{Department of Electronic Systems, Aalborg University, Denmark\\
Emails: \{maa, cs, petarp\}@es.aau.dk}

\maketitle
\begin{abstract}
We present a communication solution tailored specifically for DC microgrids (MGs) that exploits: (i) the communication potential residing in power electronic converters interfacing distributed generators to powerlines and (ii) the multiple access nature of the communication channel presented by powerlines.
The communication is achieved by modulating the parameters of the primary control loop implemented by the converters, fostering execution of the upper layer control applications.
We present the proposed solution in the context of the distributed optimal economic dispatch, where the generators periodically transmit information about their local generation capacity, and, simulatenously, using the properties of the multiple access channel, detect the aggregate generation capacity of the remote peers, with an aim of distributed computation of the optimal dispatch policy.
We evaluate the potential of the proposed solution and illustrate its inherent trade-offs.
\end{abstract}
\section{Introduction}
\label{sec:intro}

MicroGrids (MGs) are localized clusters of small-scale Distributed Energy Resources (DERs) and loads that operate either connected to the main grid or in standalone mode \cite{r1,r2}.
The MG control plane is typically organized into primary, secondary and tertiary levels \cite{r2,r3,r4}.
The primary control enables the basic operation of the system by regulating the electrical parameters (bus voltage and/or frequency) and keeping the supply-demand balance to guarantee stability.
It is implemented in a decentralized manner using the \textit{droop control law} \cite{r2,r4,r5}, relying only on local measurements.
The upper (secondary/tertiary) level control optimizes the performance of the MG in terms of maximizing the quality of the delivered power under minimal cost, and, in order to operate properly, requires exchange of information among DERs \cite{r6,r7,r8}.
An important control application is the \emph{Optimal Economic Dispatch (OED)}.
OED runs periodically, e.g., every $5-30$ minutes, and dispatches the DERs based on their generation capacities at minimum total cost.
In MGs with predominantly stochastic renewable generation, the generation capacities of the DERs vary from one dispatch period to the next and they have to be reported regularly to the OED \cite{r6}.

The traditional assumption is that an external communication system, such as wireless, covers the communication requirements of the upper level control \cite{r8}. 
However, recent works challenge this assumption due to the following issues: 1) the external communication system may jeopardize the efficiency and stability of the MG due to limited reliability and availability \cite{r5}, 2) the distributed power systems, particularly MGs, are significantly more dynamic, sporadic and ad-hoc in nature, compared to traditional centralized power system, which might deem the installation of external communication system impractical and cost inefficient \cite{r2,r5,r8,r9,r10}, and 3) making the MG system reliant on an external system contradicts the principle of self-sustainability and self-sufficiency \cite{r12,r13,r14}.
A suitable alternative is to use the existing power electronic and powerline equipment for communication \cite{r9,r10,r11,r12,r13}.
\textit{Power talk} is such solution, with a target use in direct current (DC) MGs \cite{r12,r13,r14}.
Specifically, power talk modulates information into deviations of the parameters of the primary control loops of the DERs.
In this way, a non-linear multiple access communication channel is induced, through which information-carrying deviations of the voltage (or, equivalently, power) are disseminated throughout the system, and received and processed by other DER units.
The control frequency of the primary droop controller is typically between $10-1000\,\text{Hz}$, which implies that power talk is a narrowband solution.
It exhibits similarities with other existing low-rate PLC standards for communication in the AC distribution grids, such as Ripple Carrier, TWACS and Turtle \cite{r141}, which also rely on modulating voltage to exchange information.
However, in contrast to these solutions, power talk requires no additional hardware, being implemented in the local primary control loop of the power electronic converters that connect the DERs to the DC buses.
Thus, power talk fosters the self-sustainability feature of the MG paradigm.

Previous works focused on the communication-theoretic aspects of the power talk, including the design of robust communications under variable loads, which is the major communication impairment \cite{r12,r13,r14,r15,r16,r17}.
In this paper, we use power talk to support the upper level control optimizations.
In particular, we focus on the OED and its distributed solution under linear incremental cost functions (i.e., cost per unit generation) {\cite{r6}}.
We identify the information required by the DERs to run the dispatch in distributed manner, showing that it is sufficient to locally obtain the \emph{aggregate} generation capacity of DERs with equal incremental cost.
Based on this observation, we develop a communication and computation scheme which runs periodically, in dedicated time interval prior to each dispatch period.
In the proposed scheme, DERs with equal incremental costs transmit quantized, uncoded information about their local capacities over the power talk channel in full duplex mode, whereas the receiving DERs directly detect the aggregate capacity of the transmitting DERs.
The obtained information is then used to determine the optimal dispatch in distributed manner.
The proposed solution can be viewed as a decentralized upper-level controller where the required communication capability is enabled by the primary control level that exploits the multiple-access nature of the powerlines interconnecting DERs.

The rest of the paper is organized as follows:
Section~\ref{sec:pt_mac} introduces the model of a DC MG and reviews the relevant aspects of the power talk multiple access channel.
Section~\ref{sec:doed} briefly reviews the OED and discusses its distributed solution.
Section~\ref{sec:pt_doed} presents the power talk based solution for distributed OED and Section~\ref{sec:eval} evaluates its performance.
Section~\ref{sec:conc} concludes the paper.

\section{Power Talk Multiple Access Channel}
\label{sec:pt_mac}

\begin{figure}[t]
\centering
\includegraphics[scale=0.27]{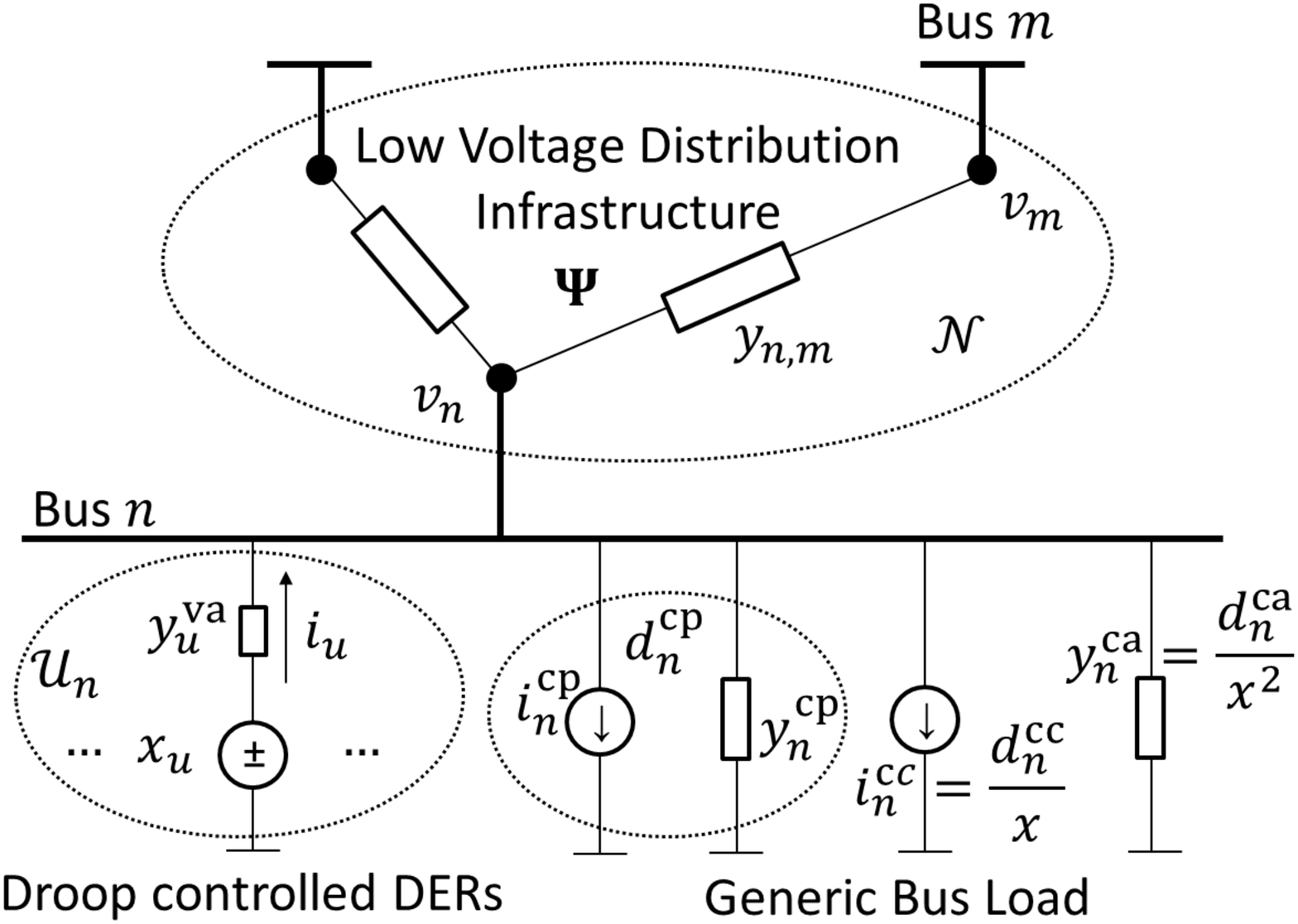}
\caption{System model: DC MicroGrid in steady state.}
\label{MBMG}
\end{figure}

\subsection{System model}

A DC MG is a collection of DERs and loads that are connected to Low Voltage DC (LVDC) distribution infrastructure, see Fig.~\ref{MBMG}.
We denote the total number of DERs with $U$, indexed in the ordered set $\mathcal{U}=\left\{0,...,U-1\right\}$.
The LVDC infrastructure consist of $N\geq 1$ buses, indexed in the ordered set $\mathcal{N}=\left\{0,...,N-1\right\}$; a \textit{bus} is defined as a point in the MG characterized by a steady state voltage denoted with $v_n,n\in\mathcal{N}$.
Bus $n\in\mathcal{N}$ hosts $U_n$ DERs, where $0< U_n <U$.
We denote the corresponding subset of DERs by $\mathcal{U}_n$, where $|\mathcal{U}_n|=U_n$ and where $\mathcal{U}_n \subset \mathcal{U}$.
We also introduce matrix {$\mathbf{E}\in\left\{0,1\right\}^{N\times U}$} with entries defined as:
\begin{align}
{e}_{n,u} & = \left\{
  \begin{array}{lr}
	   1, & u\in\mathcal{U}_n,\\
     0, & \text{otherwise}.
  \end{array}
\right.
\end{align}
The DERs use power electronic converters to interface the distribution infrastructure, and their voltage and current (i.e., power) outputs are locally regulated via primary control.
A common primary control configuration is in the form of Voltage Source Converter (VSC), see Fig.~\ref{MBMG}, which regulates the output voltage and current using the following law \cite{r2,r4,r5}:
\begin{equation}\label{eq:droop}
	v_n = x_{u} - \frac{1}{y_{u}^{\text{va}}}i_{u},\;u\in\mathcal{U},n\in\mathcal{N},
\end{equation}
where $i_{u}$ is the output current of the DER, $x_{u}$ is the \emph{reference voltage} and $y_{u}^{\text{va}}$ is the \emph{virtual admittance}.
This implementation is known as decentralized droop control.\footnote{Another primary control architecture is the Current Source Converter (CSC).
CSC units do not participate in output voltage regulation and they are operated at their generation capacity regardless of the system state, using the Maximum Power Point Tracking (MPPT) algorithm. 
%In steady state, they are modeled as constant power sources, through a negative current source and parallel admittance.
When the units engage in power talk communication, they are configured as VSC units, i.e., all CSC units, for the purpose of exchanging information via power talk switch to VSC mode of operation.}
The droop controller controls $x_{u}$ and $y_{u}^{\text{va}}$, where $x_{u}$ determines the voltage rating of the system, while the $y_{u}^{\text{va}}$ determines the power sharing among different DERs.
\footnote{In practice, the value of the virtual admittance is set to enable proportional power sharing. This aspect is discussed in more detail in Section~\ref{sec:doed}.}
%In steady state, the droop-controlled VSC units are modeled as voltage sources in series with the virtual admittance (see Fig.~\ref{MBMG}).
Besides DERs, bus $n\in\mathcal{N}$ also hosts a collection of loads, represented through an aggregate model comprising \emph{constant admittance} $y_{n}^{\text{ca}}={x^{-2}}{d_{n}^{\text{ca}}}$, \emph{constant current} $i_{n}^{\text{cc}}={x^{-1}}{d_{n}^{\text{cc}}}$, and \emph{constant power} component $d_{n}^{\text{cp}}$.
The quantities $d_{n}^{\text{ca}}$, $d_{n}^{\text{cc}}$ and $d_{n}^{\text{cp}}$ are the \emph{rated power demands} of the individual components at specified voltage $x$.
The buses are interconnected through distribution lines; the line connecting buses $n$ and $m$ is characterized by an \emph{admittance}, denoted with {$y_{n,m}\equiv y_{m,n}\geq 0$}, with equality if buses $n$ and $m$ are not directly connected, or if $n=m$.
The \emph{admittance matrix} of the system is denoted with $\mathbf{\Psi}\in\mathbb{R}^{N\times N}$, with entries defined as:
\begin{align}
{\psi}_{n,m} & = \left\{
  \begin{array}{lr}
	   \sum_{i\in\mathcal{N}}y_{n, i} & n=m,\\
     -y_{n, m} & n\neq m.
  \end{array}
\right.
\end{align}

Applying Kirchoff's laws to the system depicted in Fig.~\ref{MBMG} leads to the following current balance equation:
\begin{align}\label{eq:current_balance}
\sum_{u\in\mathcal{U}_n}(x_{u}-v_n)y_{u}^{\text{va}}= \sum_{m\in\mathcal{N}}(v_n-v_m)y_{n, m} + \frac{v_n}{x^2}d_{n}^{\text{ca}} + \frac{1}{x}d_n^{\text{cc}} + \frac{1}{v_n}d_{n}^{\text{cp}},\;n\in\mathcal{N}.
\end{align}
Solving for $v_n$, $n\in\mathcal{N}$, yields the following expression \cite{r2,r4}:
\begin{align}\nonumber
v_n & = \frac{\sum\limits_{u\in\mathcal{U}_n}x_u y_{u}^{\text{va}} + \sum\limits_{m\in\mathcal{N}}v_m y_{n, m} - \frac{d_n^{\text{cc}}}{x}}{2\bigg(\sum\limits_{u\in\mathcal{U}_n}y_{u}^{\text{va}} + \sum\limits_{m\in\mathcal{N}}y_{n, m} + \frac{d_n^{\text{ca}}}{x^2}\bigg)} \, + \\\label{eq:v_n_steady_state}
    & \frac{\sqrt{\bigg(\sum\limits_{u\in\mathcal{U}_n}x_u y_{u}^{\text{va}} + \sum\limits_{m\in\mathcal{N}}v_m y_{n, m} - \frac{d_n^{\text{cc}}}{x}\bigg)^2 - 4d_n^{\text{cp}}\bigg(\sum\limits_{u\in\mathcal{U}_n} y_{u}^{\text{va}} + \sum\limits_{m\in\mathcal{N}}y_{n, m} + \frac{d_n^{\text{ca}}}{x^2}\bigg)}}{2\bigg(\sum\limits_{u\in\mathcal{U}_n} y_{u}^{\text{va}} + \sum\limits_{m\in\mathcal{N}}y_{n, m} + \frac{d_n^{\text{ca}}}{x^2}\bigg)}.
\end{align}
In power talk, each unit $u\in\mathcal{U}$ modulates information into the values of the local reference voltage and virtual admittance droop control parameters and observes the steady state bus voltage response \cite{r12,r13,r16}.
In this regard, the model \eqref{eq:v_n_steady_state} describes the general input-output relation of the power talk \emph{multiple access channel}, where the inputs are $x_u$ and $y_{u}^{\text{va}}$, $u\in\mathcal{U}$, while the output observed by DER $u$, $u\in\mathcal{U}_n$, is $v_n$.
We comment on several important aspects: 1) the obtained multiple access channel is non-linear due to the presence of constant power loads, 2) the output is determined by the physical configuration of the system, and 3) the output is determined by the current power demand of the load components.
%From \eqref{eq:v_n_steady_state} we observe that the functional dependence of the output observed by DER $u$ on the inputs of DERs connected to other buses is algebraically implicit and solving for $v_n=v_n(x_0,...,x_{U-1},y_0^{\text{va}},...,y_{U-1}^{\text{va}}),\forall n\in\mathcal{N}$, requires knowledge of the admittance matrix and the power demands of the individual load components.
From \eqref{eq:v_n_steady_state}, we observe that solving $v_n=v_n(x_0,...,x_{U-1},y_0^{\text{va}},...,y_{U-1}^{\text{va}})$, $\forall n\in\mathcal{N}$, requires knowledge of the admittance matrix and the power demands of the individual load components.
On the primary control level, the DERs rely only on the output current as a feedback and they do not have knowledge of the MG configuration, making the power talk channel \eqref{eq:v_n_steady_state} very difficult to solve \cite{r16}.
We tackle these difficulties by using a linearized approximation of the bus voltage around a predefined operating point, which does not require detailed knowledge of the admittance matrix and the load components \cite{r14}.
 
\subsection{Discrete time linearized signal model}

We develop a linearized signal model for all-to-all full duplex power talk communication scenario, where all DERs simultaneously transmit and receive data.
We assume that the time is slotted in slots of duration $T_{\text{S}}$ and the units are slot-synchronized.\footnote{The duration of the time slot $T_{\text{S}}$ is set to comply with the control frequency of the primary controller; its value is typically of the order of milliseconds to allow the system to establish steady-state. {The aspects of achieving and maintaining slot-synchronized power talk operation is beyond the scope of the paper.} However, we note that the power electronic converters may come pre-equipped with GPS modules, providing a common time reference for achieving and maintaining slot-synchronization. {An alternative option is to use distributed algorithms for achieving synchronization, while standard line codes, e.g. Manchester code, can be used to maintain synchronization among the units}.}
In slot $t$, DER $u$ uses the following input:
\begin{align}\label{eq:input_x_n}
	x_u(t) & = \overline{x}_u+\Delta x_u(t),\\\label{eq:input_y_n}
	y_u^{\text{va}}(t) & = \overline{y}_u^{\text{va}},\;u\in\mathcal{U},
\end{align}
with $\Delta x_u(t)$ being the \textit{input} signal and $\overline{x}_u$, $\overline{y}_u^{\text{va}}$ are droop combinations defining the current operating point of the MG.
In other words, in the considered power talk variant, the information is modulated into the deviations of the reference voltage droop control parameters while the virtual admittances remain fixed.
The resulting deviation of the bus voltage in slot $t$ can be written as: 
\begin{equation}\label{eq:ch_output}
	v_n(t) = \overline{v}_n + \Delta v_{n}(t),\;n\in\mathcal{N},
\end{equation}
where $\Delta v_n(t)$ is the \textit{output} of the communication channel, and where the steady state bus voltage $\overline{v}_n$ corresponds to $\overline{x}_0,...,\overline{x}_{U-1},\overline{y}_{0}^{\text{va}},...,\overline{y}_{U-1}^{\text{va}}$. 
Unit $u$ samples the noisy version of $\Delta v_{n}(t)$ with frequency $f_{\text{S}}$ and uses the average of $N_{\text{S}}=T_{\text{S}}f_{\text{S}}$ samples over the slot $t$ to obtain the observation{\footnote{More precisely, the bus voltage is sampled after the system reaches a steady state and all transient effects diminish.}}:
\begin{align}\label{eq:y}
	\Delta \tilde{v}_{u}(t)=\Delta v_{n}(t) + z_u(t),\;u\in\mathcal{U}_n,n\in\mathcal{N},
\end{align}
where noise $z_u(t)\sim\mathcal{N}(0,\sigma^2)$ is modeled as additive Gaussian noise \cite{r16}.
Finally, we assume that the loads in the system change randomly with a rate that is {much lower than the signaling rate} $T_{\text{S}}^{-1}$ and that the signaling is done over a single realization of the load values.\footnote{Typically, the average time between consecutive load changes in MG systems is of the order of several seconds or even minutes {\cite{r6,r8}.}}

We assume that the reference voltage deviations $\Delta x_u(t)$ are relatively small:
\begin{equation}\label{lin_assump}
	\frac{|\Delta x_u(t)|}{\overline{x}_u} \ll 1\Rightarrow\frac{|\Delta v_{n}(t)|}{\overline{v}_n} \ll 1,\;u\in\mathcal{U}_n,n\in\mathcal{N}.
\end{equation}
Under this assumption, \eqref{eq:ch_output} can be linearized around $\overline{x}_0,...,\overline{x}_{U-1},\overline{y}_{0}^{\text{va}},...,\overline{y}_{U-1}^{\text{va}}$, yielding the following linear model \cite{r14}:
\begin{align}\label{eq:lin_model_vec0}
\Delta \mathbf{v}(t) \approx (\breve{\mathbf{\Psi}} + \mathbf{K}^{-1}(\mathbf{E}\mathbf{Y}^{\text{va}}\mathbf{E}^T+\mathbf{Y}^{\text{ca}}))^{-1}\mathbf{E}\mathbf{Y}^{\text{va}}\Delta\mathbf{x}(t) = \breve{\mathbf{H}}\Delta\mathbf{x}(t).
\end{align}
where $\Delta\mathbf{x}(t)=[\Delta x_u(t)]_{u\in\mathcal{U}}^T\in\mathbb{R}^{U\times 1}$, $\Delta\mathbf{v}(t)=[\Delta v_{n}(t)]_{n\in\mathcal{N}}^T\in\mathbb{R}^{N\times 1}$, $\breve{\mathbf{\Psi}}\in\mathbb{R}^{N\times N}$ is the modified admittance matrix where each diagonal entry is multiplied by $\kappa_n^{-1}$, i.e. $\breve{\psi}_{n,n}=\frac{\psi_{n,n}}{\kappa_n}$, $\mathbf{Y}^{\text{va}}=\text{diag}\left\{y_u^{\text{va}}\right\}_{u=\mathcal{U}}\in\mathbb{R}^{U\times U}$, $\mathbf{Y}^{\text{ca}}=\text{diag}\left\{y_n^{\text{ca}}\right\}_{n\in\mathcal{N}}\in\mathbb{R}^{N\times N}$, $\mathbf{K}=\text{diag}\left\{\kappa_n\right\}_{n\in\mathcal{N}}$ and $\kappa_n\geq 1$ appears as a result of linearization {(see \cite{r14})}.
We refer to matrix $\breve{\mathbf{H}}$ as the \emph{channel} matrix of the system.
The obtained linear model for the noisy output observed by DER $k$, $k\in\mathcal{U}_n$, is:
\begin{align}\label{final_signal_model1}
\Delta \tilde{v}_{k}(t)  \approx \sum_{m\in\mathcal{N}}\breve{h}_{n,m}\sum_{l\in\mathcal{U}_m}\Delta x_l(t) + z_k(t),
\end{align}
$\breve{h}_{n,m}$ is the entry at position $n,m$ of $\breve{\mathbf{H}}$; it can be shown that $\breve{h}_{n,m} > 0$, $\forall n,m$.

Due to deviations of the reference voltages, the output power of the DERs will also deviate.
Denote with $p_{u}(t)=v_n(t) i_u(t)$ the output power of DER $u\in\mathcal{U}_n$ in slot $t$.
Using assumption \eqref{lin_assump}, $p_u(t)$ can be approximated as:
\begin{align}
p_u(t) & = \overline{p}_u + \Delta p_u(t)\\\label{eq:dev_power_lin_approx}
			 & \approx \overline{p}_u + \sum_{m\in\mathcal{N}}\breve{\phi}_{u,m}\sum_{l\in\mathcal{U}_m}\Delta x_l(t),
\end{align}
where $\overline{p}_u$ corresponds to $\overline{x}_0,...,\overline{x}_{U-1},\overline{y}_{0}^{\text{va}},...,\overline{y}_{U-1}^{\text{va}}$, and where:
\begin{align}
{\breve{\phi}}_{u,m} & = \left\{
  \begin{array}{lr}
	   (\breve{h}_{n,m}\overline{x}_u - 2\breve{h}_{n,m}\overline{v}_n)\overline{y}_u^{\text{va}}, & m\neq n,\\
     (\breve{h}_{n,m}\overline{x}_u + (1-2\breve{h}_{n,m})\overline{v}_n)\overline{y}_u^{\text{va}}, & m = n.
  \end{array}
\right.
\end{align}
In practice, the amount of deviation of the output power that can be tolerated is a design parameter that constraints the input power talk signal $\Delta x_u(t)$.
%Later, we use approximation \eqref{eq:dev_power_lin_approx} to constraint the input power talk signals $\Delta x_l(t)$.
%We conclude by noting that the resulting linearized power talk communication channel \eqref{final_signal_model1} is an all-to-all full duplex Gaussian Multiple Access Channel \cite{ref:11}, provided that the channel coefficients $\breve{h}_{n,m}$, $m\in\mathcal{N}$, are known.

\section{Distributed Optimal Economic Dispatch}
\label{sec:doed}

Here we briefly review the distributed OED \cite{r6,r7,r8}.
%We are particularly interested in the information requirements imposed by the optimal solution.
From the perspective of the OED, the DERs are organized in $G$ disjoint subsets/types.
The subsets are denoted with $\mathcal{U}_g,\;g=0,...,G-1$. %see Fig.~\ref{Example} for example.
Each subset is assigned incremental cost $c_g$ per unit of generated power, where the cost $c_g$ is the same for all DERs in the subset.
Without loss of generality, assume the costs are ordered as $c_0\leq c_2\leq...\leq c_{G-1}$.
We introduce the binary matrix $\mathbf{\Xi}\in\left\{0,1\right\}^{G\times U}$, with entries defined as:
\begin{align}
{\xi}_{g,u} & = \left\{
  \begin{array}{lr}
	   1, & u\in\mathcal{U}_g,\\
     0, & \text{otherwise}.
  \end{array}
\right.
\end{align}
The generation capacity of DER $u$ at the beginning of each dispatch period is denoted with $w_{u}$.
The aggregate generation capacity of all DERs of the same type $g$ is denoted with $w^{(g)}=\sum_{u\in\mathcal{U}_g}w_u$.
The total power demand of all loads in the system during a single dispatch period is $d^{\text{L}}=\sum_{n\in\mathcal{N}}(d_n^{\text{ca}}+d_n^{\text{cc}}+d_n^{\text{cp}})$.
%The condition $\sum_{g=0}^{G-1}\sum_{u\in\mathcal{U}_g}w_u\geq d^{\text{L}}$ is satisfied for any $d^{\text{L}}$.
We assume a typical demand-response scenario where the total load demand $d^{\text{L}}$ is known \emph{a priori} (e.g., through accurate forecast programs).
In such case, the goal of the OED is to dispatch the available DER resources in optimal manner.

We define the following: 1) the \textit{power generation capacity vector} $\mathbf{w}=[w_u]_{u\in\mathcal{U}}\in\mathbb{R}^{U\times 1}$, 2) the \textit{generation cost vector} $\mathbf{c}=[c_g]_{g=0,...,G-1}\in\mathbb{R}^{G\times 1}$, and 3) the \textit{dispatch policy vector} $\mathbf{p}=[p_{u}]_{u\in\mathcal{U}}\in\mathbb{R}^{U\times 1}$.
The total power generation cost in a dispatch period is:
\begin{equation}\label{gencost}
C(\mathbf{p};\mathbf{w},\mathbf{c},d^{\text{L}})=\sum_{g=0}^{G-1}\sum_{u\in\mathcal{U}_g}c_g w_u.
\end{equation}
The \emph{optimal dispatch policy} $\mathbf{p}^{\star}$ is solution to the optimization problem:
\begin{align}\label{cOED}
               & \mathbf{p}^{\star} = \min_{\mathbf{p}}\;C(\mathbf{p};\mathbf{w},\mathbf{c},d^{\text{L}})\\\nonumber
 \text{s.t. }  & \sum_{g=0}^{G-1}\sum_{u\in\mathcal{U}_g}p_{u} = d^{\text{L}},\\\nonumber
               & 0\leq {p}_u\leq {w}_{u},u\in\mathcal{U}_g,g=0,...,G-1.
\end{align}
%The optimization problem \eqref{cOED} is linear program and can be solved efficiently by a centralized controller, provided that the vectors $\mathbf{w}$, $\mathbf{c}$ and the total power demand $d^{\text{L}}$ are available.
It can be shown that the following distributed policy is optimal for \eqref{cOED} \cite{r6}:
\begin{align}\label{dOED}
p_{u}^{\star} = \left\{
  \begin{array}{lr}
    w_{u}, & d^{\text{L}} > \sum_{j=0}^g w^{(j)}, \\
    0,     & d^{\text{L}} < \sum_{j=0}^{g-1} w^{(j)}, \\
		w_{u}\frac{(d^{\text{L}}-\sum_{j=0}^{g-1} w^{(j)})}{w^{(g)}}, & \sum_{j=0}^{g-1} w^{(j)} \leq d^{\text{L}} \leq \sum_{j=0}^g w^{(j)}.
  \end{array}
\right.
\end{align}
The first condition in \eqref{dOED} configures the DER as constant power source and inject the maximum available power into the system.
The second condition sets the unit in idle mode, i.e., the DER does not inject power into the system.
The third condition configures the DER as VSC unit for proportional power sharing, i.e., the DER employs droop control with virtual resistance set to enable proportional power sharing based on the rating $w_{u}$.

{From \eqref{dOED} it can be noted that DERs of type $g=0,...,G-1$, require the knowledge of the aggregate generation capacities $w^{(k)}$, $k \leq g$, to make the local decision, while knowledge of the generation capacities $w^{(k)}$, $k >  g$,  is not necessary.
Based on this observation, in the following section we design a power talk communication protocol to facilitate \eqref{dOED}.}

\section{Power Talk for Distributed OED}
\label{sec:pt_doed}

\begin{figure}[t]
\centering
\includegraphics[scale=0.35]{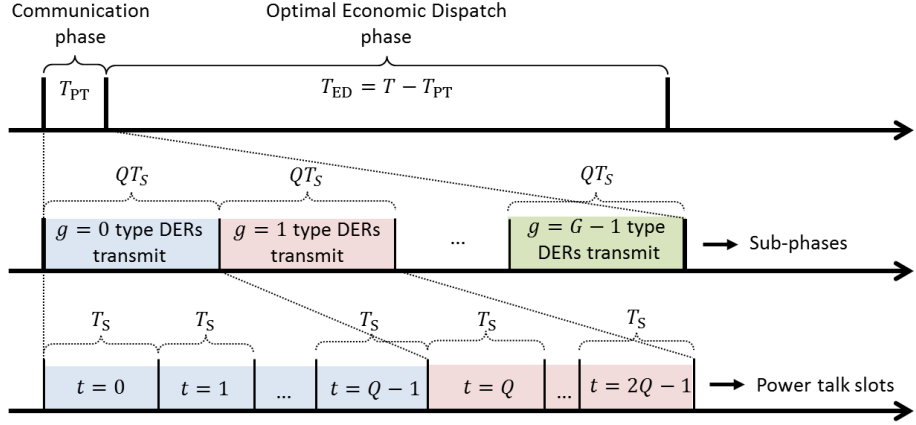}
\caption{Organization of the power talk communication protocol: phases, sub-phases and power talk slot.}
\label{TimeAxis}
\end{figure}

\subsection{Organization of the protocol operation}

Typically, economic dispatch is run periodically, with a period $T$ ranging from $5$ up to $30$ minutes.
%The duration of the economic dispatch period is denoted with $T$, .\footnote{}
We split this period into two phases, see Fig.~\ref{TimeAxis}: 1) \emph{communication phase} of duration $T_{\text{PT}}$, in which the units exchange information via the power talk multiple access channel \eqref{final_signal_model1}, and 2) \emph{OED phase} of {duration $T_{\text{ED}}=T-T_{\text{PT}}$, in which the configuration of each DER is determined by the outcome of the optimal decentralized algorithm \eqref{dOED}, based on  the information obtained during the communication phase.}\footnote{We note that the power talk communication in practice should also involve a channel estimation phase, where the DERs estimate the coefficients $\breve{h}_{n,m}$; {this aspect is out of the paper scope and we assume that the channel coefficients are known (see \cite{r16} for a detailed discussion on the channel estimation in power talk)}.}
The communication phase is split in $G$ sub-phases of duration $T_{g}$, $g=0,...,G-1$, and each sub-phase is divided into $Q$ time slots of duration $T_S$, such that $T_g = Q T_{\text{S}}$, $\forall g$.
We assume that each time slot is indexed with index $ i = gQ+t$, where $t=0,...,Q-1$ and $g=0,...,G-1$.
%The scheduling of the transmitting and receiving DERs in sub-phase $g$ is determined as follows:
The communication is organized on the sub-phase basis, as follows:
\begin{itemize}
\item All DERs of type $g$ simultaneously transmit in sub-phase $g$, i.e., there are $|\mathcal{U}_g|$ transmitters in sub-phase $g$;
\item All DERs of types $g,g+1,...,G-1$ receive in sub-phase $g$; i.e., there are $\sum_{j=g}^{G-1}|\mathcal{U}_j|$ receivers in sub-phase $g$.
\end{itemize}
Obviously, DERs of type $g$ work in full duplex mode in sub-phase $g$.
By the end of sub-phase $g$, the DERs of type $g$ will collect all aggregate generation capacities $w^{(0)},...,w^{(g)}$ that are required to run the decentralized OED \eqref{dOED}.
%Notice that the transmitting DERs of type $g$ also act as receivers, i.e. they operate in full duplex mode and listen to transmissions of other DERs from the same type.

When transmitting, DER $u$ transmits a bit sequence (i.e., a word) that represents quantized value of its local generation capacity $w_u$, denoted by $\check{w}_u$.
On the other hand, the receiving units do not resolve the individual $\check{w}_{u}$ for each $u\in\mathcal{U}_g$ that transmits simultaneously; rather, they detect the sum ${w}^{(g)}$ directly from the observations, as described below.

\subsection{Detecting aggregate generation capacities}

Each transmitting DER $u\in\mathcal{U}_g$ quantizes the local value $w_{u}$ with step $\bigtriangleup$ and $2^Q$ quantization levels, using the following rule:
\begin{equation}\label{quant}
\text{if }w_{u}\in[\beta_u\bigtriangleup,(\beta_u+1)\bigtriangleup)\Rightarrow \check{w}_{u}=\bigg(\beta_u+\frac{1}{2}\bigg)\bigtriangleup.
\end{equation}
The quantization index $\beta_u\in\left\{0,...,2^Q-1\right\}$ of the transmitting DER $u$ is represented with uncoded stream of $Q$ bits:
\begin{equation}\label{eq:bit_stream}
\left\{b_u(gQ+t)\in\left\{0,1\right\}\right\}_{t=0,...,Q-1},\;u\in\mathcal{U}_g.
\end{equation}
Each bit $b_u(gQ+t)$ is then mapped into a corresponding power talk input (i.e., deviation of the reference voltage):
\begin{align}\label{eq:antipodal_sig}
 \Delta x_u(gQ+t) & = \left\{
  \begin{array}{lr}
	   \lambda, & b_u(gQ+t)=1,\\
     -\lambda, & b_u(gQ+t)=0.
  \end{array}
\right.
\end{align}
A receiving DER receiving in slot $gQ+t$ of sub-phase $g$ obtains the following measurement:
\begin{align}
{\Delta{\tilde{v}}_k^{'}(gQ+t)} & = \Delta \tilde{v}_{k}(gQ+t) - \breve{h}_{n,n}\xi_{g,k}\Delta x_k(gQ+t)\\
& = \sum_{m\in\mathcal{N}} \breve{h}_{n,m}\sum_{\substack{l\in\mathcal{U}_m,l\neq k}}\xi_{g,l}\Delta x_{l}(gQ+t) + z_k(gQ+t)\\
										& = \lambda\sum_{m\in\mathcal{N}} \breve{h}_{n,m}\sum_{\substack{l\in\mathcal{U}_m,l\neq k}}\xi_{g,l}\big(2b_l(gQ+t)-1\big) + z_k(gQ+t).
\end{align}
Combining the measurements $\Delta{\tilde{v}}_k^{'}(gQ+t)$, $t=0,...,Q-1$, collected in sub-phase $g$, the goal of the receiving DER $k$ is to detect:
\begin{align}
\check{w}^{(g)} & = \sum_{u\in\mathcal{U}_g,u\neq k}\check{w}_{u} = \bigg(\sum_{u\in\mathcal{U}_g,u\neq k}\beta_u + \frac{|\mathcal{U}_g|-\xi_{g,k}}{2}\bigg)\bigtriangleup \\\label{eq:new1}
								& = \bigg(\sum_{t=0}^{Q-1}\sum_{u\in\mathcal{U}_g,u\neq k}b_u(gQ+t)2^{t}+\frac{|\mathcal{U}_g|-\xi_{g,k}}{2}\bigg)\bigtriangleup.
\end{align}
Thus, the receiving DER $k$ needs to determine the integer sum $\sum_{u\in\mathcal{U}_g}b_u(gQ+t)=\theta^{(g)}(gQ+t)$ of the bits received in slot $gQ+t$.
The optimal Maximum-A-Posteriori (MAP) detection of $\theta^{(g)}(gQ+t)$ in slot $gQ+t$, is defined as follows:
\begin{equation}\label{MAP_gen}
\hat{\theta}^{(g)}(gQ+t) = \max_{\theta}f(\Delta{\tilde{v}}_k^{'}(gQ+t);{\theta}^{(g)}(gQ+t)=\theta)\text{Pr}({\theta}^{(g)}(gQ+t)=\theta),
\end{equation}
where $f(\Delta{\tilde{v}}_k^{'}(gQ+t);{\theta}^{(g)}(gQ+t)=\theta)$ is the density function of the measurement $\Delta{\tilde{v}}_k^{'}(gQ+t)$, parametrized w.r.t. the integer sum $\theta=0,...,|\mathcal{U}_g|-\xi_{g,k}$, and $\text{Pr}({\theta}^{(g)}(gQ+t)=\theta)$ is the \emph{a priori} probability of $\theta$.
Note that the maximization is performed w.r.t. $\theta = 0,...,|\mathcal{U}_g|-\xi_{g,k}$, which implies that the complexity of the detector grows linearly with the number of simultaneously transmitting DERs in a sub-phase. 
Under Gaussian noise assumption, \eqref{MAP_gen} becomes:
\begin{equation}\label{MAP}
\hat{\theta}^{(g)} = \max_{\theta}\sum_{j=1}^{\binom{|\mathcal{U}_g|-\xi_{g,k}}{\theta}}\exp{\left\{-\frac{(\Delta{\tilde{v}}_k^{'}-\lambda\sum_{m\in\mathcal{N}} \breve{h}_{n,m}\sum_{\substack{l\in\mathcal{U}_m,l\neq k}}\xi_{g,l}(2b_l^j-1))^2}{2\sigma^2}\right\}},
\end{equation}
where the slot index $gQ+t$ is omitted due to space limitation.
The summation in \eqref{MAP}  is over all $\binom{|\mathcal{U}_g|-\xi_{g,k}}{\theta}$ binary sequences $b_l^j(gQ+t)$, $j=0,...,\binom{|\mathcal{U}_g|-\xi_{g,k}}{\theta}-1$ of length $|\mathcal{U}_g|-\xi_{g,k}$.
These sums are computed by each receiving DER and stored in memory for each $\theta=0,...,|\mathcal{U}_g|-\xi_{g,k}$ and each $g$.
Using \eqref{MAP}, a DER receiving in sub-phase $g$ detects the aggregate generation capacity:
\begin{align}\label{eq:gen_est}
\hat{\check{w}}^{(g)} = \bigg(\sum_{t=0}^{Q-1}\hat{\theta}^{(g)}(gQ+t)2^{t}+\frac{|\mathcal{U}_g|-\xi_{g,k}}{2}\bigg)\bigtriangleup.
\end{align}
In the case when $G=U$, i.e., the costs per unit output power are different for all DERs, the above power talk strategy reduces to simple TDMA solution, where a single DER transmits in each sub-phase.

Finally, we provide a policy for choosing the parameter $\lambda$.
We constraint the variance of the output power deviations of each DER as follows:
\begin{equation}
\text{Var}(\Delta p_u)\leq \pi^2,\;u\in\mathcal{U},
\end{equation}
where $\pi$ is the power deviation budget of each unit, corresponding to the system tolerance to power deviations.
This constraint yields the following range for $\lambda$:
\begin{equation}
0 < \lambda\leq \min_{u}\left\{\frac{\pi}{\sqrt{\sum_{m\in\mathcal{N}}\breve{\phi}_{u,m}^2\sum_{l\in\mathcal{U}_m}\xi_{g,l}}}\right\},
\end{equation}
where we used the linear approximation \eqref{eq:dev_power_lin_approx} for $\Delta p_u$.

\subsection{Cost trade-off}

At the end of the communication phase, each DER $u\in\mathcal{U}$ operates with imperfect knowledge of the sum generation capacities $w^{(g)}$, due to quantization and detection error.
The corresponding, potentially suboptimal dispatch policy vector obtained via \eqref{dOED} is denoted with $\hat{\check{\mathbf{p}}}^{\star}$.
The total generation cost when the policy {$\hat{\check{\mathbf{p}}}^{\star}$} is enforced, is denoted with $\hat{\check{C}}^{\star}$.
Further, a dispatch policy might under- or over-estimate the cumulative generation capacities, leading to power deficit $p^{\text{def}}=(d^{\text{L}}-\sum_{g=0}^{G-1}\sum_{u\in\mathcal{U}_g}\hat{\check{p}}_u^{\star})^{+}$ or power surplus $p^{\text{sur}}=(\sum_{g=0}^{G-1}\sum_{u\in\mathcal{U}_g}\hat{\check{p}}_u^{\star}-d^{\text{L}})^{+}$. %\footnote{By definition, the function $f(x)=(x)^{+}=x$ if $x>0$ and 0 otherwise.}.
In the case of deficit, a back-up source (e.g., a storage system or supply from the main grid) is activated; the cost per unit generation of the back-up source is denoted with $c^{\text{def}}$.
Similarly, the power surplus is transferred to storage system/main grid at cost $c^{\text{sur}}>0$ \cite{r6,r8}.

The total generation cost for a dispatch period can be calculated as follows:
\begin{align}
C(\hat{\check{\mathbf{p}}}^{\star}) & = \bigg(1-\frac{T_{\text{S}}}{T}QG\bigg)\underbrace{(\hat{\check{C}}^{\star} + \varrho)}_{\hat{\check{\Omega}}^{\star}} + \frac{T_\text{S}}{T}\sum_{g=0}^{G-1}\sum_{t=0}^{Q-1}\sum_{g=0}^{G-1}\sum_{u\in\mathcal{U}_g}c_gp_u(gQ+t)\\\label{cost12}
& = \hat{\check{\Omega}}^{\star} + \frac{T_{\text{S}}}{T}QG\bigg(\sum_{g=0}^{G-1}\sum_{u\in\mathcal{U}_g}c_g(\overline{p}_u - \hat{\check{p}}_u^{\star}) -\varrho\bigg),
\end{align}
where $\varrho = c^{\text{def}}p^{\text{def}} + c^{\text{sur}}p^{\text{sur}}$, and where we assumed that $p_u(gQ+t)=\overline{p}+\Delta p_u(gQ+t)$ and $\lim_{Q\rightarrow\infty}\sum_{t=0}^{Q-1}\Delta p_u(gQ+t) = 0$, i.e., it is assumed that antipodal signaling, see \eqref{eq:antipodal_sig}, results in (roughly) symmetric supply power deviations around $\overline{p}_u$.
Eq. \eqref{cost12} provides insight into the fundamental trade-off of the proposed solution.
Namely, $\lim_{Q\rightarrow\infty,T_{\text{S}}\rightarrow\infty}\hat{\check{\Omega}}^{\star}={{C}}^{\star}<\hat{\check{\Omega}}^{\star}$; however, increasing $Q$ and/or the slot duration, increases the duration of the communication phase where the system operates suboptimally, %mode, at the expense of the duration of the dispatch phase where the system operates in (quasi)optimal model,
potentially increasing the overall cost.

\section{Evaluation}
\label{sec:eval}

In this section, we evaluate the performance of the proposed technique by simulating a single bus system (i.e., $N=1$), to which all DERs and loads are connected through lines with negligible resistances.
This is a valid model for small, localized MGs, where the effect of the transmission network on the power flow is negligible \cite{r6,r8}.
There are $U=10$ DERs, organized into $G=4$ types:
\begin{equation}
\mathbf{\Xi} = 
\begin{bmatrix}
1 & 1 & 1 & 0 & 0 & 0 & 0 & 0 & 0 & 0\\
0 & 0 & 0 & 1 & 1 & 0 & 0 & 0 & 0 & 0\\
0 & 0 & 0 & 0 & 0 & 1 & 1 & 1 & 0 & 0\\
0 & 0 & 0 & 0 & 0 & 0 & 0 & 0 & 1 & 1
\end{bmatrix}.
\end{equation}
The cost vector is $\mathbf{c}=[5,5,5,7.5,7.5,10,10,10,50,50]$ and $c^{\text{def}}=c^{\text{sur}}=100$; note that these numbers are used only for illustrative purposes.
The power generation of each DG changes uniformly and independently in the interval $w_{u}\in[0,w_{\max}=2\,\text{kW}]$, while the total load power demand is $d^{\text{L}}=5\,\text{kW}$; the load is composed only of constant power part, i.e., $d^{\text{L}}=d^{\text{cp}}$.
The quantization step is $\bigtriangleup=\frac{w_{\max}}{2^Q}$.
The total duration of the dispatch period is fixed to $T=300\,\text{s}$.
The sampling frequency of the converter's front-end is $f_{\text{S}}=50\;\text{kHz}$, and the standard deviation of the voltage sampling noise is $\sqrt{N_0}=0.1\;\text{V}/\text{sample}$ \cite{r16}.
We investigate the cost behavior as a function of the number of bits $Q$ for varying slot durations $T_{\text{S}}$ and different tolerances $\pi$ on the output power deviations.

\begin{figure}[t]
\centering
\includegraphics[scale=0.4]{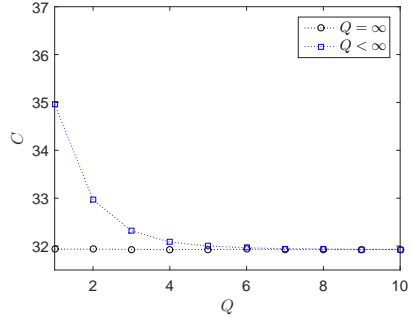}
\caption{Optimal economic dispatch: effect of quantization error.}
\label{Res1}
\end{figure}
\begin{figure}[t]
\centering
\includegraphics[scale=0.4]{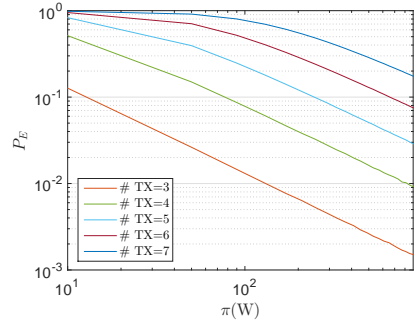}
\caption{Error probability of the detector \eqref{MAP}, parametrized by the number of simultaneously transmitting DERs.}
\label{Res34}
\end{figure}

First, we illustrate the effect of the quantization error on the optimal decentralized dispatch policy \eqref{dOED}, presented in Fig.~\ref{Res1}.
%We assume perfect and instantaneous underlying communication network to support the dispatch application.
It can be noted that this effect becomes negligible for $Q>10$.
This implies that in usual MG control applications, the length of the messages that need to be exchanged among the units is exceptionally short.
In fact, in virtually all upper level control applications, less than $2\;\text{bytes}$ of information per node message is sufficient \cite{r6}.

Next, we investigate the performance of the detector \eqref{MAP}.
The detector operates with averages over single power talk slot, i.e., the observed value $\Delta \tilde{v}_u(gQ+t)$ is obtained by averaging $T_{\text{S}}f_{\text{S}}$ samples during the slot $gQ+t$.
Therefore, the standard deviation of the noise component $z_u(gQ+t)\sim\mathcal{N}(0,\sigma^2)$ in each slot, is $\sigma = \sqrt{\frac{N_0}{T_{\text{S}}f_{\text{S}}}}$.
%Evidently, the slot duration impacts the overall observation noise: increasing $T_{\text{S}}$ decreases the noise deviation (i.e., noise power).
%However, increasing $T_{\text{S}}$ also increases the duration of the communication phase at the expense of the duration of the dispatch phase, leading to a trade-off.
\begin{figure*}[t]
\centering
\subfloat[$T_S=10\, \text{ms}$]{\includegraphics[scale=0.4]{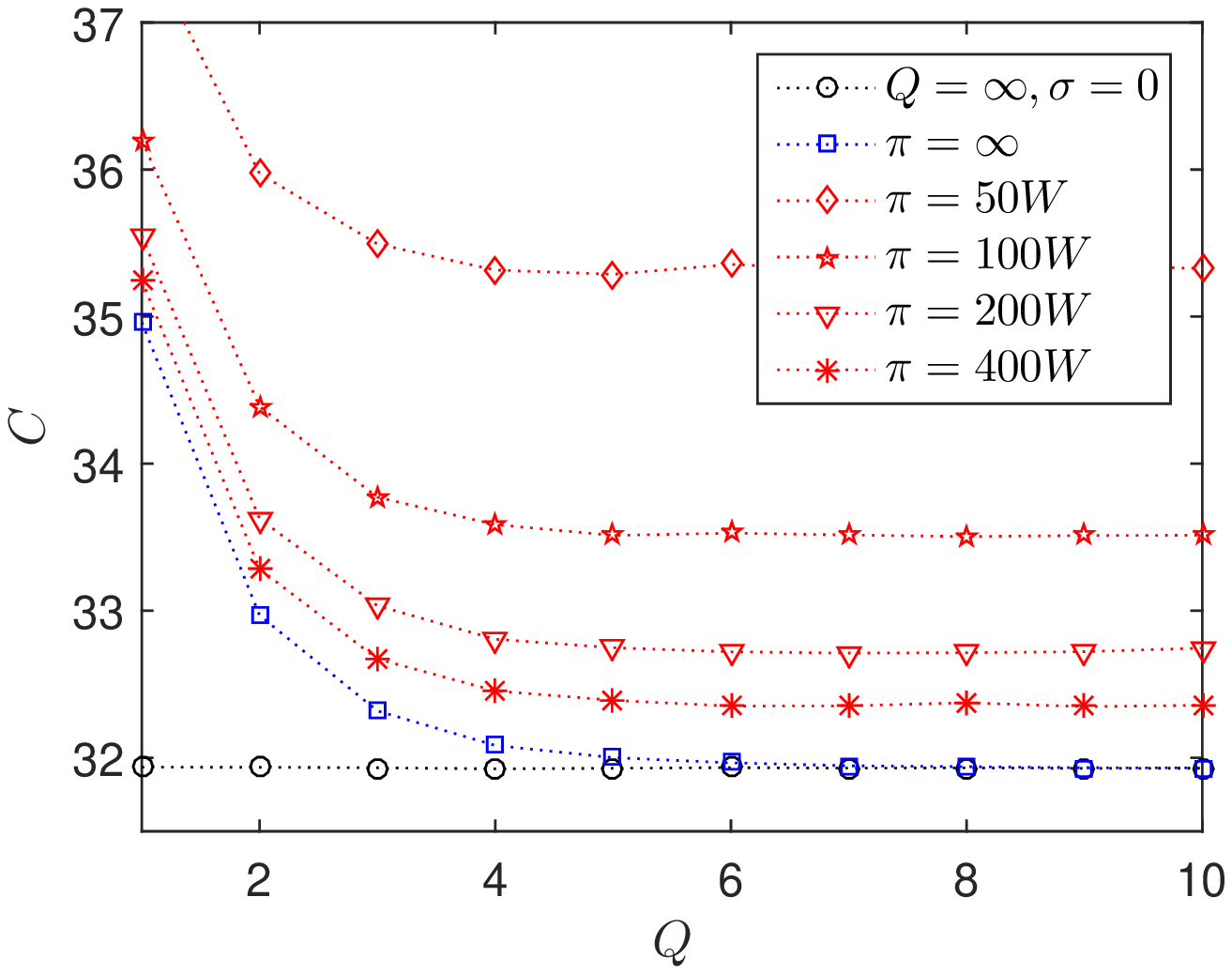} \label{a}}
\hfil
\subfloat[$T_S=50\, \text{ms}$]{\includegraphics[scale=0.4]{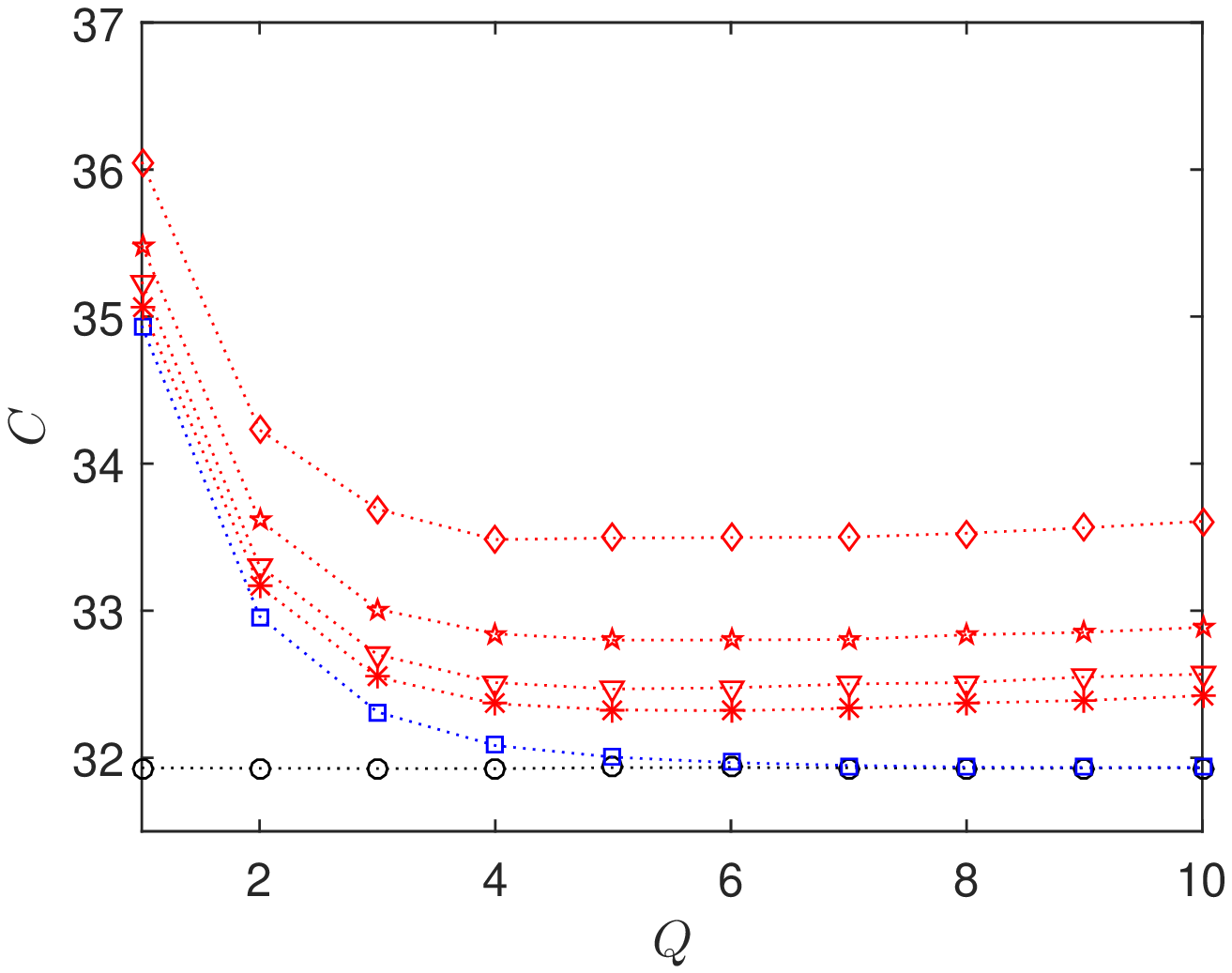} \label{b}}
\hfil
\subfloat[$T_S=150\, \text{ms}$]{\includegraphics[scale=0.4]{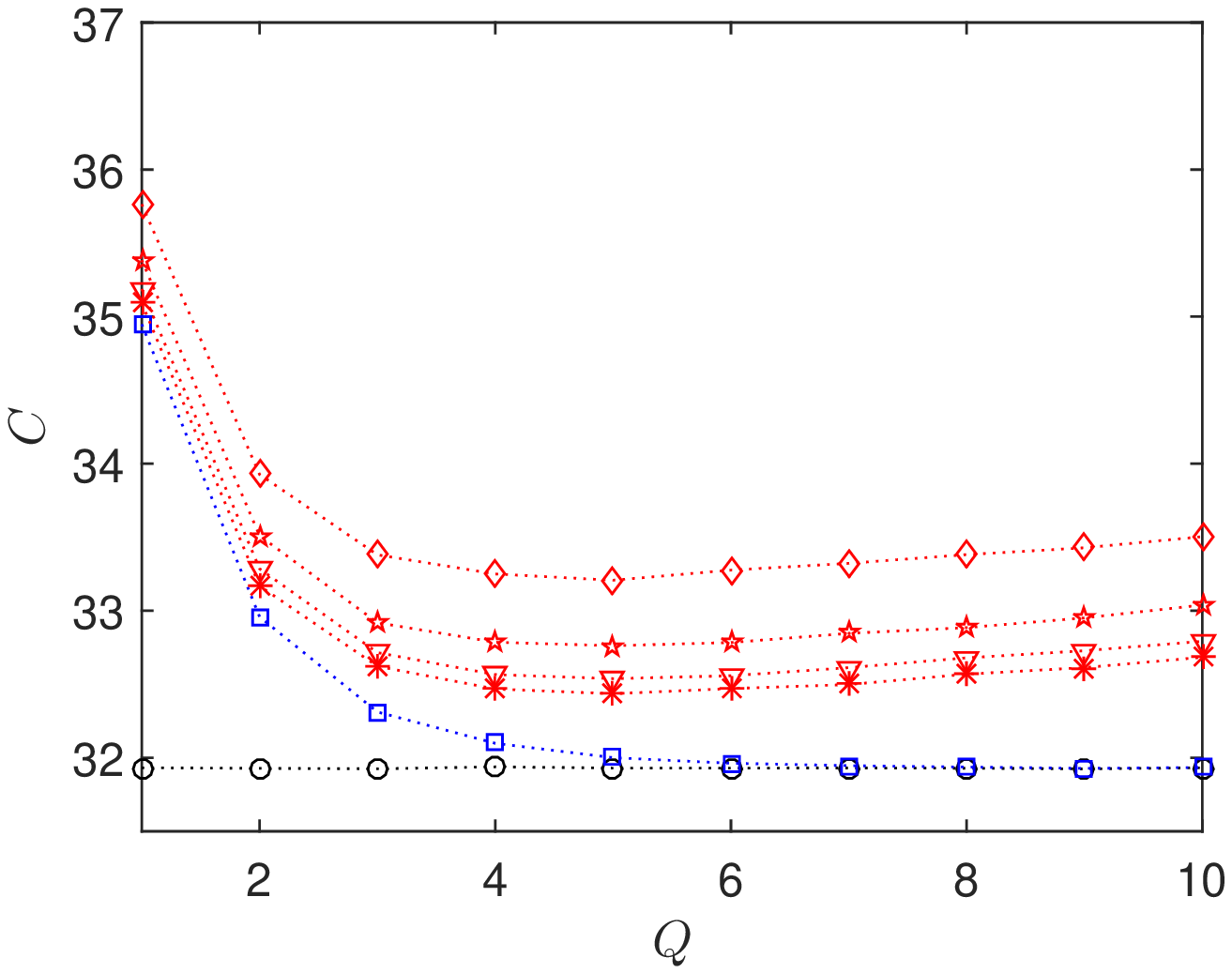} \label{c}}
\hfil
\subfloat[$T_S=200\, \text{ms}$]{\includegraphics[scale=0.4]{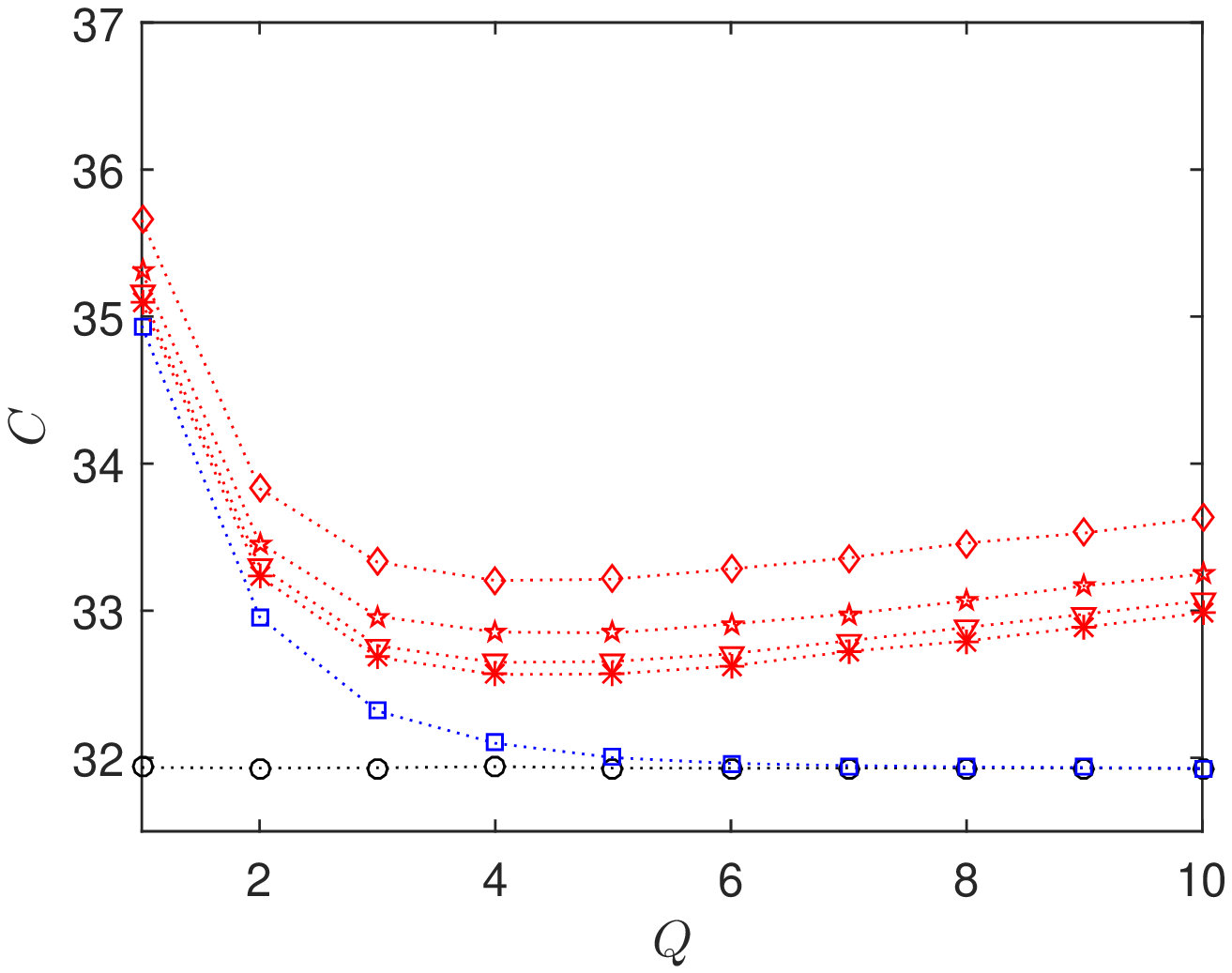} \label{d}}
\caption{Generation cost of the optimal decentralized dispatch policy $\hat{\check{p}}^{\star}$.}
\label{Res5}
\end{figure*}
{Fig.~\ref{Res34} shows the average probability of error $P_E$ of the detector \eqref{MAP} using $T_{\text{S}}=0.1\;\text{s}$ and $Q=10$} as a function of $\pi$, for increasing number of simultaneously transmitting units.
Obviously, the probability of error increases as the number of units increases.
On the other hand, applications that can tolerate larger output voltage deviations, i.e., larger $\pi$, can benefit from improved detector performance.
It can be concluded that our detector is well suited for MG control applications as in typical MG setup the total number of DERs is low, typically less than $\sim 10$.

Fig.~\ref{Res5} illustrates the total generation cost of the dispatch policy $\hat{\check{p}}^{\star}$ \eqref{cost12}.
Increasing $T_{\text{S}}$ suppresses the noise (recall that $\sigma = \sqrt{\frac{N_0}{T_{\text{S}}f_{\text{S}}}}$), which improves the detector performance and pushes the first term in \eqref{cost12} towards the optimal value.
At the same time, the overall duration of the power talk phase is increased, which increases the value of the second term.
One way to improve the first term in \eqref{cost12} while keeping the second term fixed is through increasing $\pi$; however, the choice of $\pi$ will be determined by the specific amount of deviation of the electric parameters that can be tolerated in the system.
For instance, in LVDC MG system with rated voltage of $400\,\text{V}$, reference voltage deviations of $\pm 2\,\text{V}$ amount to $\pm 0.5\%$ deviation from the the rated value; these deviations preserve the small signal assumption \eqref{lin_assump} and allow values for $\pi$ of up to $200\, \text{W}$, which significantly improves the performance of the detector, see Fig.~\ref{Res34}.
We conclude that power talk indeed shows strong potential as a communication enabler for upper layer control applications in MGs.
%An additional important advantage stems from the fact that power talk is a solution implemented over the already existing primary control level within the power electronic converters, requires no additional hardware and it has the same reliability and availability as the MG system it self.

\section{Concluding Remarks}
\label{sec:conc}

We presented a powerline communication protocol for control applications in DC MicroGrids, specifically designed to facilitate distributed optimal economic dispatch without support of an external communication network.
In the proposed solution, the distributed generators, transmit information about their local, instantaneous generation capacities over the power lines in full duplex mode, while the receiving generators use specific integer sum detector to retrieve the aggregate generation capacity of the transmitting generators.
On the physical layer, the solution exploits the multiple access nature of the power talk communication channel in which, information is modulated into the parameters of the primary control loops of power electronic converters.
The simulation results illustrate the inherent trade-offs and prove that the propose solution is a viable communication alternative for self-sustainable and self-sufficient DC MG systems.

}\setcounter{section}{\value{save}}

\end{document}